\documentclass[aps,prl,twocolumn,groupedaddress]{revtex4}
\usepackage[usenames,dvipsnames]{color}

\usepackage{graphicx}
\usepackage{verbatim}

\newcommand{\suppress}[1]{}
\newcommand{\Jensen}{Gibbs}
\newcommand{\beq}{\begin{eqnarray}}
\newcommand{\eeq}{\end{eqnarray}}

\newcommand{\beginfig}{\begin{figure}}
\newcommand{\fref}[1]{Fig. \ref{#1}}
\newcommand{\eref}[1]{Eqn. \ref{#1}}
\newcommand{\vect}[1]{\ensuremath{\vec{#1}}}
\newcommand{\mA}{\mathbf{A}}
\newcommand{\Aij}{A_{ij}}
\newcommand{\thetap}{\theta_+}
\newcommand{\thetam}{\theta_-}

\newcommand{\vtheta}{\vect{\theta}}
\newcommand{\vpi}{\vect{\pi}}

\newcommand{\sumij}{\sum_{i>j}}

\newcommand{\intpi}{\int \!\! d\vpi}
\newcommand{\inttheta}{\int \!\! d\vtheta}

\newcommand{\ptheta}{p(\vtheta)}

\newcommand{\ppi}{p(\vpi)}

\newcommand{\npp}{n_{++}}
\newcommand{\npm}{n_{+-}}
\newcommand{\nmp}{n_{-+}}
\newcommand{\nmm}{n_{--}}

\newcommand{\sumi}{\sum_{i=1}^N}

\newcommand{\am}{\alpha_-}
\newcommand{\bp}{\beta_+}
\newcommand{\bm}{\beta_-}
\newcommand{\apt}{\tilde{\alpha}_+}
\newcommand{\amt}{\tilde{\alpha}_-}
\newcommand{\bpt}{\tilde{\beta}_+}
\newcommand{\bmt}{\tilde{\beta}_-}

\newcommand{\expval}[1]{\langle #1 \rangle}
\newcommand{\FQA}{F\{Q;\mA\}}

\newcommand{\mQ}{\mathbf{Q}}

\newcommand{\Qimu}{Q_{i\mu}}
\newcommand{\Qjmu}{Q_{j\mu}}

\newcommand{\enpp}{\expval{\npp}}
\newcommand{\enpm}{\expval{\npm}}
\newcommand{\enmp}{\expval{\nmp}}
\newcommand{\enmm}{\expval{\nmm}}
\newcommand{\enmu}{\expval{\nmu}}

\newcommand{\va}{\vect{\alpha}}
\newcommand{\prodmu}{\prod_{\mu=1}^K}

\newcommand{\summu}{\sum_{\mu=1}^K}
\newcommand{\pimu}{\pi_{\mu}}

\newcommand{\nmu}{n_{\mu}}
\newcommand{\amu}{\alpha_{\mu}}

\newcommand{\amut}{\tilde{\amu}}

\newcommand{\Z}{\mathcal{Z}}
\newcommand{\Zp}{\Z_+}
\newcommand{\Zm}{\Z_-}
\newcommand{\Zpi}{\Z_{\vpi}}
\newcommand{\Zpt}{\tilde{\Z}_+}
\newcommand{\Zmt}{\tilde{\Z}_-}
\newcommand{\Zpit}{\tilde{\Z}_{\vpi}}

\newcommand{\hamiltonian}{\mathcal{H}}

\newcommand{\Hspin}{\hamiltonian_{\spin|\param}}

\newcommand{\deltasisj}{\delta_{\sigma_i,\sigma_j}}
\newcommand{\deltasimu}{\delta_{\sigma_i,\mu}}
\newcommand{\hmu}{h_{\mu}}

\renewcommand{\npp}{c_+}
\renewcommand{\npm}{c_-}
\renewcommand{\nmp}{d_+}
\renewcommand{\nmm}{d_-}
\renewcommand{\thetap}{\theta_c}
\renewcommand{\thetam}{\theta_d}
\renewcommand{\ap}{\tilde{c}_+}

\renewcommand{\apt}{\tilde{c}_{+_0}}
\renewcommand{\bp}{\tilde{c}_-}

\renewcommand{\bpt}{\tilde{c}_{-_0}}
\renewcommand{\am}{\tilde{d}_+}
\renewcommand{\amt}{\tilde{d}_{+_0}}
\renewcommand{\bm}{\tilde{d}_-}
\renewcommand{\bmt}{\tilde{d}_{-_0}}
\renewcommand{\va}{\tilde{\vect{n}}}
\newcommand{\vn}{\vect{n}}
\renewcommand{\amu}{\tilde{n}_{\mu}}
\renewcommand{\amut}{\tilde{n}_{\mu_0}}
\newcommand{\si}{\sigma_i}
\newcommand{\sj}{\sigma_j}
\newcommand{\vs}{\vect{\sigma}}
\newcommand{\sums}{\sum_{\vs}}

\newcommand{\sumlocal}{\sumij \Aij \deltasisj}

\newcommand{\sumnpp}{\sumlocal}
\newcommand{\sumnpm}{\sumij (1-\Aij) \deltasisj}
\newcommand{\sumnmp}{\sumij \Aij (1-\deltasisj)}
\newcommand{\sumnmm}{\sumij (1-\Aij) (1-\deltasisj)}

\renewcommand{\vtheta}{\vect{\theta}}
\newcommand{\pAspit}{p(\mA,\vs,\vpi,\vtheta|K)}
\newcommand{\pAgspit}{p(\mA|\vs,\vtheta)}
\newcommand{\pAsgpitK}{p(\mA,\vs|\vpi,\vtheta,K)}
\newcommand{\pAsgpit}{p(\mA,\vs|\vpi,\vtheta)}
\newcommand{\psgpi}{p(\vs|\vpi)}
\newcommand{\ppitgA}{p(\vpi,\vtheta|\mA)}
\newcommand{\psgA}{p(\vs|\mA)}
\newcommand{\pAgK}{p(\mA|K)}
\newcommand{\qspit}{q(\vs,\vpi,\vtheta)}
\renewcommand{\FQA}{F\{q;\mA\}}
\newcommand{\qs}{q_{\vs}(\vs)}
\newcommand{\qpi}{q_{\vpi}(\vpi)}
\newcommand{\qt}{q_{\vtheta}(\vtheta)}
\newcommand{\qc}{q_c(\thetap)}
\newcommand{\qd}{q_d(\thetam)}

\renewcommand{\Zp}{\Z_c}
\renewcommand{\Zm}{\Z_d}
\renewcommand{\Zpt}{\tilde{\Z}_c}
\renewcommand{\Zmt}{\tilde{\Z}_d}
\renewcommand{\Hspin}{\hamiltonian}

\renewcommand{\theta}{\vartheta}

\begin{document}

\title{A Bayesian Approach to Network Modularity}
\author{Jake M. Hofman}
 \email{jmh2045@columbia.edu}
 \affiliation{Department of Physics, Columbia University, New York, NY 10027}
\author{Chris H. Wiggins}
 \email{chris.wiggins@columbia.edu}
 \affiliation{Department of Applied Physics and Applied Mathematics, Columbia University, New York, NY 10027}

\date{\today}

\begin{abstract}
We present an efficient, principled, and interpretable technique for
inferring module assignments and for identifying the optimal number of
modules in a given network. We show how several existing methods for
finding modules can be described as variant, special, or limiting
cases of our work, and how the method overcomes the resolution limit
problem, accurately recovering the true number of modules.  Our
approach is based on Bayesian methods for model selection which have
been used with success for almost a century, implemented using a
variational technique developed only in the past decade.  We apply the
technique to synthetic and real networks and outline how the method
naturally allows selection among competing models.
\end{abstract}

\pacs{}

\maketitle

Large-scale networks describing complex interactions among a multitude
of objects have found application in a wide array of fields, from
biology to social science to information technology
\cite{RevModPhys.74.47,watts98}.  In these applications one often wishes
to {\it model} networks, suppressing the complexity of the full
description while retaining relevant information about the structure
of the interactions \cite{Ziv:2005aa}. One such network model groups
nodes into modules, or ``communities,'' with different densities of
intra- and inter- connectivity for nodes in the same or different
modules.  We present here a computationally efficient Bayesian
framework for inferring the number of modules, model parameters, and
module assignments for such a model.

The problem of finding modules in networks (or ``community
detection'') has received much attention in the physics literature,
wherein many approaches \cite{newman:026113,2006PhRvE..74a6110R} focus
on optimizing an energy-based cost function with fixed parameters over
possible assignments of nodes into modules. The particular cost
functions vary, but most compare a given node partitioning to an
implicit null model, the two most popular being the configuration
model and a limited version of the stochastic block model (SBM)
\cite{holland,mcsherry}. While much effort has gone into {\it how} to
optimize these cost functions, less attention has been paid to {\it
  what} is to be optimized. In recent studies which emphasize the
importance of the latter question it was shown that there are inherent
problems with existing approaches {\it regardless of how optimization
  is performed}, wherein parameter choice sets a lower limit on the
size of detected modules, referred to as the ``resolution limit''
problem \cite{fortunato,kumpula}. We extend recent probabilistic
treatments of modular networks
\cite{2006PhRvE..74c5102H,Newman:2007aa} to develop a solution to this
problem that relies on {\it inferring} distributions over the model
parameters, as opposed to {\it asserting} parameter values {\it a
  priori}, to determine the modular structure of a given network.  The
developed techniques are principled, interpretable, computationally
efficient, and can be shown to generalize several previous studies on
module detection.

We specify an $N$-node network by its adjacency matrix $\mA$, where
$\Aij=1$ if there is an edge between nodes $i$ and $j$ and $\Aij=0$
otherwise, and define $\si\in\{1,\ldots,K\}$ to be the unobserved
module membership of the $i^{\rm th}$ node.  We use a constrained SBM,
which consists of a multinomial distribution over module assignments
with weights $\pimu \equiv p(\si=\mu|\vpi)$ and Bernoulli
distributions over edges contained within and between modules with
weights $\thetap \equiv p(\Aij=1|\si=\sj,\vtheta)$ and $\thetam \equiv
p(\Aij=1|\si \neq \sj,\vtheta)$, respectively. In short, to generate a
random undirected graph under this model we roll a $K$-sided die
(biased by $\vpi$) $N$ times to determine module assignments for each
of the $N$ nodes; we then flip one of two biased coins (for either
intra- or inter- module connection, biased by $\thetap, \thetam$,
respectively) for each of the $N(N-1)/2$ pairs of nodes to determine
if the pair is connected. The extension to directed graphs is
straightforward.

Using this model, we write the joint probability
$\pAsgpitK=\pAgspit\psgpi$ (conditional dependence on $K$ has
been suppressed below for brevity) as
\begin{equation}
    \pAsgpit =  \thetap^{\npp} (1-\thetap)^{\npm} \thetam^{\nmp} (1-\thetam)^{\nmm} \prodmu \pimu^{\nmu}
\end{equation}
where $\npp\equiv\sumnpp$ is the number of edges contained within
communities, $\npm\equiv\sumnpm$ is the number of non-edges contained
within communities, $\nmp\equiv\sumnmp$ is the number of edges between
different communities, $\nmm\equiv\sumnmm$ is the number of non-edges
between different communities, and $\nmu\equiv\sumi \delta_{\si,\mu}$
is the occupation number of the $\mu^{\rm th}$ module. Defining
$\Hspin \equiv -\ln \pAsgpit$ and regrouping terms by local and global
counts, we recover (up to additive constants) a generalized version of
\cite{2006PhRvE..74c5102H}:
\begin{equation}
  \Hspin = -\sumij \left( J_L \Aij  - J_G \right) \deltasisj + \summu \hmu \sumi \deltasimu,
\end{equation}
a Potts model Hamiltonian with unknown coupling constants $J_G\equiv
\ln (1-\thetam)/(1-\thetap)$, $J_L \equiv \ln {\thetap/\thetam}+J_G$,
and chemical potentials $\hmu \equiv -\ln\pimu$. (Note that many
previous methods omit a chemical potential term, implicitly assuming
equally-sized groups.)

While previous approaches
\cite{2006PhRvE..74a6110R,2006PhRvE..74c5102H} minimize related
Hamiltonians as a function of $\vs$, these methods require that the
user specifies values for these unknown constants, 
which gives rise to the resolution limit problem
\cite{fortunato,kumpula}. Our approach, however, uses a
disorder-averaged calculation to infer distributions over these
parameters, avoiding this issue. To do so, we take beta $(\mathcal{B})$ and
Dirichlet $(\mathcal{D})$ distributions over $\vtheta$ and $\vpi$,
respectively:
\begin{equation}
    \ptheta \ppi \equiv \mathcal{B}(\thetap;\apt,\bpt)
    \mathcal{B}(\thetam;\amt,\bmt) \mathcal{D}(\vpi;\va_0).
\end{equation}
These {\it conjugate prior} distributions, are defined on the full
range of $\vtheta$ and $\vpi$, respectively, and their functional
forms are preserved when integrated against the model to obtain
updated parameter distributions. Their hyperparameters
$\{\apt,\bpt,\amt,\bmt,\va_0\}$ act as {\it pseudocounts} that augment
observed edge counts and occupation numbers.

In this framework the problem of module detection can be stated as
follows: given an adjacency matrix $\mA$, determine the most probable
number of modules (i.e. occupied spin states) $K^*=\mathrm{argmax}_K
\ p(K|\mA)$ and infer posterior distributions
over the model parameters (i.e. coupling constants and chemical
potentials) $\ppitgA$ and the latent module assignments (i.e. spin
states) $\psgA$. In the absence of {\it a priori} belief about the
number of modules, we demand that $p(K)$ is sufficiently weak that
maximizing $p(K|\mA) \propto \pAgK p(K)$ is equivalent to maximizing
$\pAgK$, referred to as the {\it evidence}. This approach to model
selection \cite{kass95bayes} proposed by the statistical physicist
Jeffreys in 1935 \cite{Jeffreys:35} balances model fidelity and
complexity to determine, in this context, the number of modules.

A more physically intuitive interpretation of the evidence is as the
disorder-averaged partition function of a spin-glass, calculated by
marginalizing over the possible quenched values of the parameters
$\vtheta$ and $\vpi$ as well as the spin configurations $\vs$:
\begin{eqnarray}
\label{eqn:evidence} 
  \Z = \pAgK &=& \sums \inttheta \intpi \ \pAsgpit \ptheta \ppi  \\
        &=& \sums \inttheta \intpi \ e^{-\Hspin} \ptheta \ppi.
\end{eqnarray}

While the $\vtheta$ and $\vpi$ integrals in \eref{eqn:evidence} can be
performed analytically, the remaining sum over module assignments
$\vs$ scales as $K^N$ and becomes computationally intractable for
networks of even modest sizes. To accommodate large-scale networks we
use a variational approach that is well-known to the statistical
physics community \cite{feynman} and has recently found application in
the statistics and machine learning literature, commonly termed
variational Bayes (VB) \cite{jordan99introduction}.  We proceed by
taking the negative logarithm of $\Z$ and using \Jensen's inequality:
\begin{eqnarray}
-\ln\Z &=&-\ln \sums \inttheta \intpi \ \qspit {\pAspit \over \qspit} \\
&\le& -\sums \inttheta \intpi \ \qspit \ln {\pAspit \over \qspit}.
\label{eqn:FAQ}
\end{eqnarray}
That is, we first multiply and divide by an arbitrary approximating
distribution $\qspit$ and then upper-bound the log of the expectation
by the expectation of the log. We define the quantity to be minimized
-- the expression in \eref{eqn:FAQ} -- as the variational free energy
$\FQA$, a functional of $\qspit$. (Note that the negative log of
$\qspit$ plays the role of a test Hamiltonian in variational
approaches in statistical mechanics.)

We next choose a factorized approximating distribution $\qspit=
\qs\qpi\qt$ with $\qpi=\mathcal{D}(\vpi;\vn)$ and
$\qt=\qc\qd=\mathcal{B}(\thetap;\ap,\bp)\mathcal{B}(\thetam;\am,\bm)$;
as in mean field theory, we factorize $\qs$ as $q(\si=\mu)=\Qimu$, an
$N$-by-$K$ matrix which gives the probability that the $i$-th node
belongs to the $\mu$-th module.  Evaluating $\FQA$ with this
functional form for $\qspit$ gives a function of the variational
parameters $\{\ap,\bp,\am,\bm,\va\}$ and matrix elements $\Qimu$ which
can subsequently be minimized by taking the appropriate derivatives.

We summarize the resulting iterative algorithm, which provably
converges to a local minimum of $\FQA$ and provides controlled
approximations to the evidence $\pAgK$ as well as the posteriors
$\ppitgA$ and $\psgA$:


{\it Initialization.}---Initialize the $N$-by-$K$ matrix $\mQ=\mQ_0$ and set pseudocounts $\ap=\apt,\bp=\bpt,\am=\amt,\bm=\bmt,$ and $\amu=\amut$.

{\it Main Loop.}---Until convergence in $\FQA$:

(i) Update the expected value of the coupling constants and chemical
potentials
    \begin{eqnarray}
      \expval{J_L} &=& \psi(\ap)-\psi(\bp)-\psi(\am)+\psi(\bm) \\
      \expval{J_G} 
         &=& \psi(\bm)-\psi(\am+\bm)\nonumber\\
         && -\psi(\bp)+\psi(\ap+\bp) \\
      \expval{\hmu}&=&\psi\left(\sum_{\mu}\amu\right)-\psi(\amu),
    \end{eqnarray}
where $\psi(x)$ is the digamma function;

(ii) Update the variational distribution over each spin $\si$
    \begin{equation}
      \Qimu \propto \exp \left\{\sum_{j \neq i} 
                         \left[\expval{J_L}\Aij-\expval{J_G}\right] \Qjmu
                         -\expval{\hmu}\right\}
    \end{equation}
normalized such that $\sum_{\mu}Q_{i\mu}=1$, for all $i$;

(iii) Update the variational distribution over parameters from the
expected counts and pseudocounts
    \begin{eqnarray}
      \amu &=& \enmu+\amut=\sumi \Qimu + \amut \\
      \ap &=& \enpp+\apt={1\over2}Tr(\mQ^T\mA\mQ) + \apt\\
      \bp &=& \enpm+\bpt \nonumber \\
      &=&{1\over2}Tr(\mQ^T(\vect{u}\expval{\vn}^T-\mQ)) -\enpp + \bpt\\
      \am &=& \enmp+\amt=M-\enpp + \amt\\
      \bm &=& \enmm+\bmt=C-M-\enpm + \bmt,
    \end{eqnarray}
where $C=N(N-1)/2$, $M=\sumij \Aij$, and $\vect{u}$ is a $N$-by-$1$
vector of 1's;

(iv) Calculate the updated optimized free energy
    \begin{eqnarray}
	\FQA &=& -\ln {\Zp \Zm \Zpi \over \Zpt \Zmt \Zpit} 
	+\summu \sumi \Qimu \ln \Qimu,
    \end{eqnarray}
where $\Zpi={B}(\va)$ is the beta function with a vector-valued
argument, the partition function for the Dirichlet distribution $\qpi$
(likewise for $\qc,\qd$).

As this provably converges to a local optimum, VB is
best implemented with multiple randomly-chosen initializations of
$\mQ_0$ to find the global minimum of $\FQA$.  

Convergence of the above algorithm provides the approximate posterior
distributions $\qs, \qpi$, and $\qt$ and simultaneously returns $K^*$,
the number of non-empty modules that maximizes the evidence.  As such,
one needs only to specify a maximum number of allowed modules and run
VB; the probability of occupation for extraneous modules converges to
zero as the algorithm runs and the most probable number of occupied
modules remains.

\begin{figure}
\begin{center}
\label{fig:reslim_vb}\includegraphics[width=.2\textwidth]{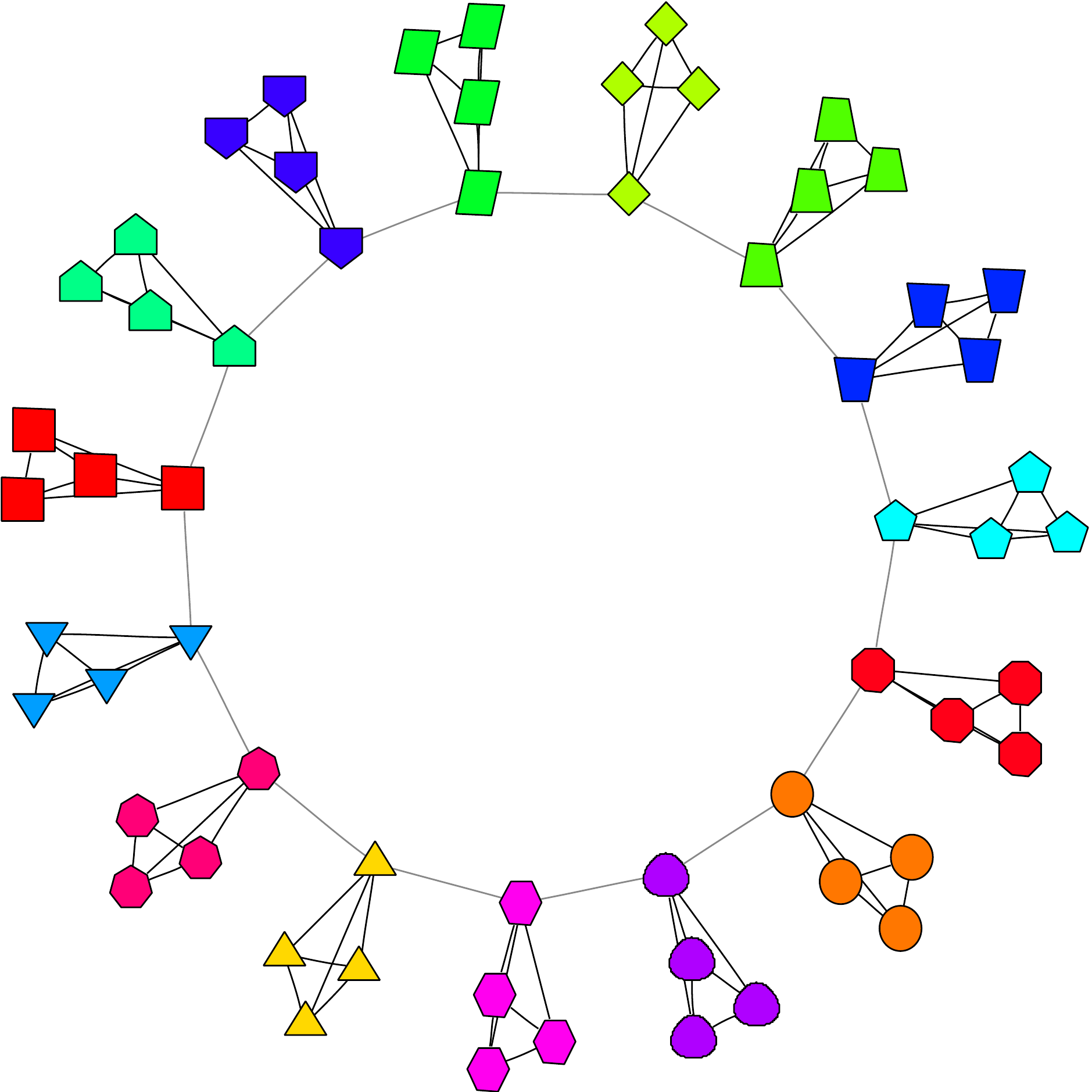}
\label{fig:reslim_mod}\includegraphics[width=.2\textwidth]{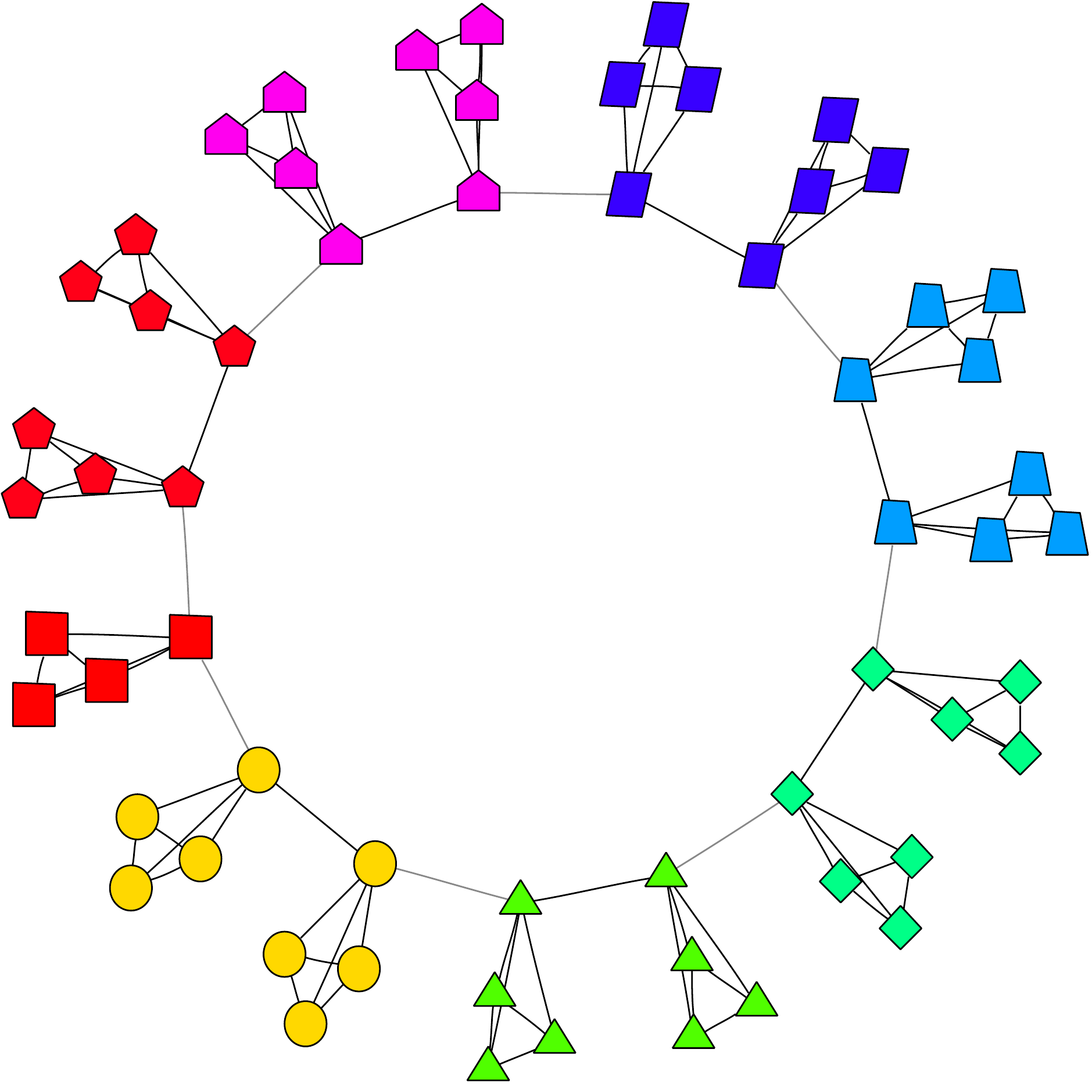}
\label{fig:reslim}
      \begin{tabular}{r|ccccccccccc}
         ${K_{\rm True}}$ & 10 & 11 & 12 & 13 & 14 & 15 & 16 & 17 & 18 & 19 & 20 \\
          \hline 
         ${K_{\rm VB}}$ & \bf{10} & \bf{11} & \bf{12} & \bf{13} & \bf{14} & \bf{15} & \bf{16} & \bf{17} & \bf{18} & \bf{19} & \bf{20} \\
         ${K_{\rm GN}}$ & \bf{10} & \bf{11} & \bf{12} & \bf{13} & \bf{14} & 8 & 9 & 9 & 10 & 11 & 11
      \end{tabular}
\caption{Results for the resolution limit test suggested in
  \cite{fortunato} and \cite{kumpula}. Shapes and colors correspond to
  the inferred modules. (Left) Our method, variational Bayes, in which
  all 15 modules are correctly identified (each clique is assigned a
  unique color/shape). (Right) GN modularity optimization, where
  failure due to the resolution limit is observed -- neighboring
  cliques are incorrectly grouped together.  (Bottom) The results of
  this test implemented for a range of true number of modules, $K_{\rm
    true}$, the number of 4-node cliques in the ring-like graph. Note
  that our method correctly infers the number of communities $K_{\rm
    VB}$ over the entire range of $K_{\rm True}$, while GN modularity
  initially finds the correct number of communities but fails for
  $K_{\rm True} \ge 15$ as shown analytically in \cite{fortunato}.}
\end{center}
\end{figure}

This is significantly more accurate than other approximate methods,
such as Bayesian Information Criterion (BIC) \cite{schwarz} and
Integrated Classification Likelihood (ICL)
\cite{ICL_orig,ICL_network}, and is less computationally expensive
than empirical methods such as cross-validation (CV) \cite{stone,xing}
in which one must perform the associated procedure after fitting the
model for each considered value of $K$. Specifically, BIC and ICL are
suggested for single-peaked likelihood functions well-approximated by
Laplace integration and studied in the large-$N$ limit. For a SBM the
first assumption of a single-peaked function is invalidated by the
underlying symmetries of the latent variables, i.e. nodes are
distinguishable and modules indistinguishable. See \fref{fig:reslim}
for comparison of our method with the Girvan-Newman modularity
\cite{newman:026113} in the resolution limit test
\cite{fortunato,kumpula}, where VB consistently identifies the correct
number of modules. (Note that VB is both accurate and fast: it
performs competitively in the ``four groups'' test \cite{danon} and
scales as $\mathcal{O}(MK)$. Runtime for the main loop in
MATLAB on a 2GHz laptop is $\sim\!\!6$ minutes for $N=10^6$ nodes with average degree 16 and $K=4$.)

Furthermore, we note that previous methods in which parameter inference
is performed by optimizing a likelihood function via Expectation
Maximization (EM) \cite{Newman:2007aa,ICL_network} are also 
special cases of the framework presented here. EM is a limiting case
of VB in which one collapses the distributions over parameters to
point-estimates at the mode of each distribution; however EM is prone
to overfitting and cannot be used to determine the appropriate number
of modules, as the likelihood of observed data increases
with the number of modules in the model. As such, VB performs at least
as well as EM while simultaneously providing complexity control
\cite{bishop,mackay}.

\begin{figure}
\begin{center}
\includegraphics[width=.44\textwidth]{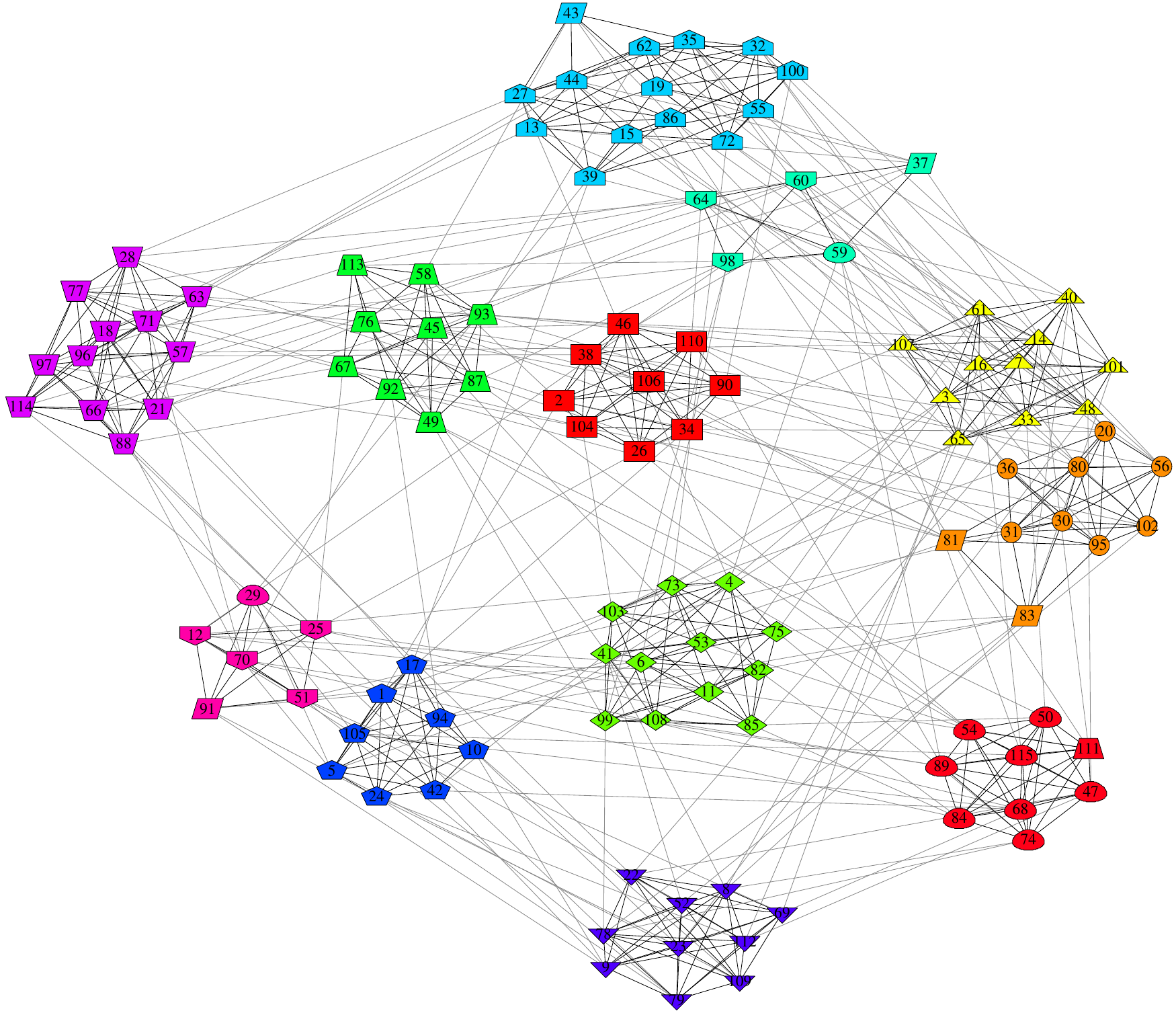}

\caption{Each of the 115 nodes represents a NCAA team and each of the
613 edges a game played in 2000 between two teams it joins. The
inferred module assignments (designated by color) on the football
network which recover the 12 NCAA conferences (designated by
shape). Nodes 29, 43, 59, 60, 64, 81, 83, 91, 98, and 111 are
misclassified and are mostly independent teams,
represented by parallelograms.}

\label{fig:football}
\end{center}
\end{figure}

In addition to validating the method on synthetic networks, we apply
VB to the 2000 NCAA American football schedule shown in
\fref{fig:football} \cite{Girvan:2002aa}. Each of the 115 nodes
represents an individual team and each of the 613 edges represents a
game played between the nodes joined. The algorithm correctly
identifies the presence of the 12 conferences which comprise the
schedule, where teams tend to play more games within than between
conferences, making most modules assortative. Of the 115 teams, 105
teams are assigned to their corresponding conferences, with the
majority of exceptions belonging to the frequently-misclassified
independent teams \cite{clauset:icml} -- the only disassortative
group in the network. We emphasize that, unlike other methods in
which the number of conferences must be asserted, VB determines 12 as
the most probable number of conferences automatically.

Posing module detection as inference of a latent variable within a
probabilistic model has a number of advantages. It clarifies what
precisely is to be optimized and suggests a principled and efficient
procedure for how to perform this optimization. Inferring
distributions over model parameters reveals the natural scale of a
given modular network, avoiding resolution limit problems. This method
allows us to view a number of approaches to the problem by physicists,
applied mathematicians, social scientists, and computer scientists as
related subparts of a larger problem. In short, it suggests how a
number of seemingly-disparate methods may be re-cast and united. A
second advantage of this work is its generalization to other models,
including those designed to reveal structural features other than
modularity. Finally, use of the evidence allows model selection not
only among nested models, e.g. models differing only in the number of
parameters, but even among models of different parametric
families. The last strikes us as a natural area for progress in the
statistical study of real-world networks.

It is a pleasure to acknowledge useful conversations on modeling with
Joel Bader and Matthew Hastings, on Monte Carlo methods for Potts
models with Jonathan Goodman, with David Blei on variational methods,
and with Aaron Clauset for his feedback on this manuscript.  J.H. was
supported by NIH 5PN2EY016586; C.W. was supported by NSF ECS-0425850
and NIH 1U54CA121852.

\bibliographystyle{apsrev}
\bibliography{genmod_prl_nourl}
\end{document}